\documentclass{mem}
\usepackage{natbib}\usepackage{txfonts}\usepackage{balance}
\usepackage{graphicx}
\usepackage[a4paper,breaklinks,dvipdfm]{hyperref}
\idline{75}{282}

\begin{document}
\def\ltsim{\raise 2pt \hbox {$<$} \kern-0.6em \lower 3pt \hbox {$\sim$}}
\def\ltapprox{\raise 2pt \hbox {$<$} \kern-1.1em \lower 5pt \hbox {$\approx
$}}
\def\gtsim{\raise 2pt \hbox {$>$} \kern-1.1em \lower 4pt \hbox {$\sim$}}

\title {Diffuse radio sources in colliding galaxy clusters
}

   \subtitle{Low frequency follow up of the GMRT Radio Halo Survey}

\author{
S. \,Giacintucci\inst{1,2,3} 
          }

  \offprints{S. Giacintucci}

\institute{
Department of Astronomy, University of Maryland, College Park, MD 20742-2421
\and
Harvard-Smithsonian Center for Astrophyisics, 60 Garden St.,
Cambridge, MA 02138
\and
INAF-IRA, Via Gobetti 101, I-40129 Bologna, Italy
\\
\email{simona@astro.umd.edu}
}

\authorrunning{Giacintucci}

\titlerunning{Diffuse radio sources in colliding galaxy clusters}

\abstract{The knowledge of the origin and statistical properties of 
diffuse radio emission in galaxy clusters has appreciably improved thanks 
to the {\em GMRT} Radio Halo Survey, a project based on 610 MHz 
observations of clusters belonging to a statistically complete sample. 
However, the spectral properties of cluster diffuse sources are still 
poorly known and uncertain. High sensitivity and multi-resolution 
observations at low frequency ($\le$0.3 GHz) are needed for accurate
spectral studies. Here,  {\em GMRT} images at 325 MHz are presented 
for the clusters A\,2744, A\,1300, A\,1758N and A\,781, all hosting
cluster-scale diffuse emission in the form of a giant halo and/or relic.
These observations are part of a new observational campaign to 
follow up with the {\em GMRT} at 150, 235 and 325 MHz all diffuse 
radio sources in the cluster sample of the {\em GMRT} Radio Halo
Survey and obtain detailed information on their radio spectral properties.
}
\maketitle{}

\section{Introduction}

The presence of non-thermal phenomena in galaxy clusters is 
nowadays firmly assessed by radio observations. Synchrotron giant
radio halos and peripheral relics prove the existence of
relativistic particles and magnetic fields in clusters
\citep{ferrari08, cassano09}  -- see also Venturi, Brunetti and 
Murgia, this volume.

Radio relics are believed to arise at merger shocks via
diffusive Fermi acceleration \citep{en98,hoeft07,weeren10}. 
A spatial connection between an X-ray shock and radio emission is 
indeed observed in a number of clusters (e.g., Markevitch 2010 and
this volume). While relics mark the locations of shocks, 
giant halos may be caused by in-situ reacceleration of pre-existing, 
lower-energy and long-lived relativistic electrons by turbulence 
during cluster mergers \citep{brunetti01,petrosian01}. Hadronic 
collisions between thermal and relativistic protons may be alternative 
or additional sources of relativistic electrons (\cite{pfrommer08};  
Pfrommer, this volume).

The 610 MHz {\em GMRT} Radio Halo Survey, investigating a statistically 
complete sample of clusters \citep{venturi07,venturi08}, addressed 
the question whether all clusters exhibit a giant radio halo. The
survey led to the detection of new diffuse cluster-scale sources and 
provided a statistical confirmation that giant halos are hosted
exclusively by massive and merging systems \citep{brunetti07,cassano08, cassano10} -- see also Cassano, this volume.

Here, we present some initial results of a low frequency follow up
with the {\em GMRT} at 325, 235 and 150 MHz 
of all halos, relics and candidates in the 
{\em GMRT} Radio Halo Survey sample. The project aims to 
increase our knowledge of the spectral properties 
of halos and relics, which is still very poor and uncertain. 
Most of them have been imaged only at 1.4 GHz 
and interferometric high-sensitivity observations at low frequency 
 ($\le$ 0.3 GHz), with the range of resolutions needed to perform 
such study, are available only for a handful of clusters.
\\
\\
We use H$_0$=70 km/s/Mpc, $\Omega_m$=0.3,  $\Omega_{\Lambda}$=0.7.

\section{GMRT observations and images}

Halo and relic emission usually embeds a number of discrete sources,
which need to be thoroughly removed to perform proper
imaging and accurate flux density measurements. The wide range of
spacings provided by the {\em GMRT} configuration is particularly well
suited for removal of discrete sources, as it provides a simultaneous
range of resolutions, which allow to accurately subtract individual
objects and image the diffuse emission in its whole extent.

The sample selected for the low frequency follow up of the {\em GMRT} Radio
Halo Survey includes all clusters with a detected radio halo or relic 
and those with candidate diffuse emission. Table 1 provides the list
of clusters observed so far. Three clusters have been subject of
dedicated papers (see notes to the table); the 235 and 325 MHz
observations of the remaining ones will be presented in
\cite{venturi11a}. The reduction of the 150 MHz data is in 
progress (Macario, this volume). 

Here, we present the 325 MHz images of A\,2744, A\,781, A\,1300 and 
A\,1758N. These clusters were observed with the {\em GMRT} for a total 
of $\sim$8 hours each, using the default spectral-line mode, with a 
32 MHz total band. The data were calibrated and reduced as described 
in \cite{giacintucci08}. Residual amplitude errors are $\ltsim$ 5\%.  

Despite the massive data editing required by 
radio frequency interference, the quality of the 325 MHz
images is good, with a 1$\sigma$ rms level in the
range $0.1{-}0.5$ mJy/beam at full resolution ($\sim$
10$^{\prime \prime}$).  Sets of low resolution (35$^{\prime\prime}$-
40$^{\prime\prime}$) images were also produced for each cluster, 
after careful subtraction of the individual radio sources from the 
u-v data. 

\begin{table}
\caption{Clusters observed with the GMRT}
\label{abun}
\begin{center}
\begin{tabular}{lcc}
\hline
\\
Cluster & $\nu$ & Source \\
  & (MHz) &             \\
\hline
\\
$*$ A\,2744  & 325 & R+H \\
\phantom{00}A\,209    & 325 & H \\
\phantom{00}A\,521   &  150, 235,325 & \phantom{0}R+H $^{1,2}$\\
\phantom{00}A\,697   & 150, 325 & \phantom{00}H $^{2,3}$\\
$*$ A\,781 & 325 & \phantom{0}R $^{4}$ \\
\phantom{00}Z\,2661 & 325 &cH \\
$*$ A\,1300 & 325 & R+H\\
\phantom{00}A\,1682 & 150, 235 & \phantom{0}cH $^{2}$\\
\phantom{00}RXCJ1314.4-2515 & 235,325 & 2R+H \\
$*$ A1758N & 325& H \\
\phantom{00}RXCJ2003.5-2323 & 235,325 & \phantom{0}H $^5$ \\
\\
\hline
\end{tabular}
\end{center}
Notes to the Table.  R= relic, H=halo, cH=candidate halo, cR=candidate
relic; *: clusters
presented here; 1: \cite{giacintucci08}, \cite{brunetti08},
\cite{dallacasa09}; 2: Macario, this volume; 3: \cite{macario10};
4: \cite{venturi11b}; 5: \cite{giacintucci09}.
\end{table}

\begin{figure}[t!]
\centering
\includegraphics[scale=0.34]{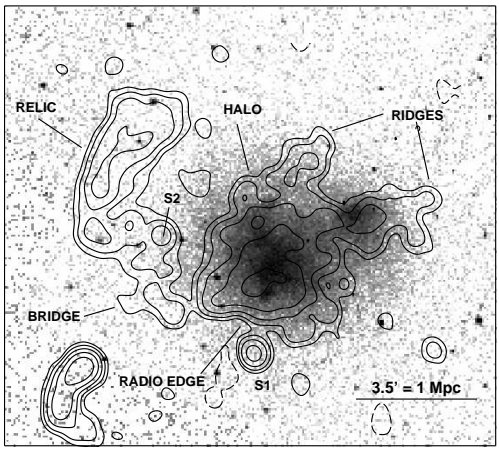}
\caption{\footnotesize {\bf A\,2744}. 
{\em GMRT} 325 MHz contours overlaid on the X-ray {\em Chandra} image in the
0.5-4.0 kev band. The radio beam is 35$^{\prime\prime}\times$35$^{\prime\prime}$.
Contours start at +3$\sigma=$1 mJy/beam and then scale by a factor of
2. The $-3\sigma$ level is shown as dashed contours. S1 and S2
indicate the position of discrete radio galaxies.  
}
\label{fig:a2744}
\end{figure}

\subsection{A\,2744} 

A\,2744 is a spectacular example of cluster-cluster merger, likely
in a post-core passage state, complicated by the infall of a third 
subcluster from the NW \citep{owers11}. An X-ray shock front
was recently discovered SE of the main cluster \citep{markevitch10, owers11}.
The cluster is impressive also in the radio band, hosting a very
large radio halo and a giant relic \citep{govoni01, orru07}. 

Fig.~\ref{fig:a2744} shows the {\em GMRT} 325 MHz low-resolution contours
on the {\em Chandra} image. The halo and relic appear connected by a
faint bridge, visible also at 1.4 GHz \citep{govoni01}. 
The radio halo is larger than previously imaged at 327~MHz
\citep{orru07}; including the two prominent ridges of emission 
detected toward the N and NW, the halo reaches a size of 
$\sim$1.9 Mpc. The relic has a higher surface brightness compared to 
the halo, and its size is $\sim1.3$ Mpc, in agreement with the
literature images. The bridge extends for $\sim$700 kpc between the
halo and relic. 

The halo covers the whole cluster X-ray emission. Interestingly, the
NW ridge seems to follow the same direction of the subcluster X-ray
emission. A positional shift between the X-ray and radio peaks is
observed for both the main cluster and subcluster. The relic is
located just outside the detected X-ray emission. The shock front in
the SE periphery of the cluster coincides with an edge-like feature of
the radio halo, as observed in few other radio-halo clusters 
(Markevitch 2010 and this volume).

The halo has a flux density of 323$\pm$16 mJy at 325 MHz
(excluding the bridge and S1), and the relic has $S_{\rm 325~MHz}=122\pm6$
mJy (bridge and S2 excluded). The flux density in the bridge is $\sim~30$
mJy. These measurements are considerably higher than those in Orr\'u
et al. (2007). Integration of the {\em GMRT} image over the same area
imaged by Orr\'u et al. gives $S_{\rm 325~MHz \, GMRT} = 280\pm14$ mJy, to
be compared to $S_{\rm 325~MHz \, VLA}=218\pm10$ mJy. For this reason, 
the resulting 0.3-1.4 GHz spectral index is steeper than reported in
\cite{giovannini09}: $\alpha=1.2/1.1 $ (using $S_{\rm 325~MHz \, GMRT)}
=323/280$ mJy) is found for the halo and $\alpha$=1.3 for the relic. 

\subsection{A\,1300} 

\begin{figure*}[t!]
\centering
\includegraphics[scale=0.61]{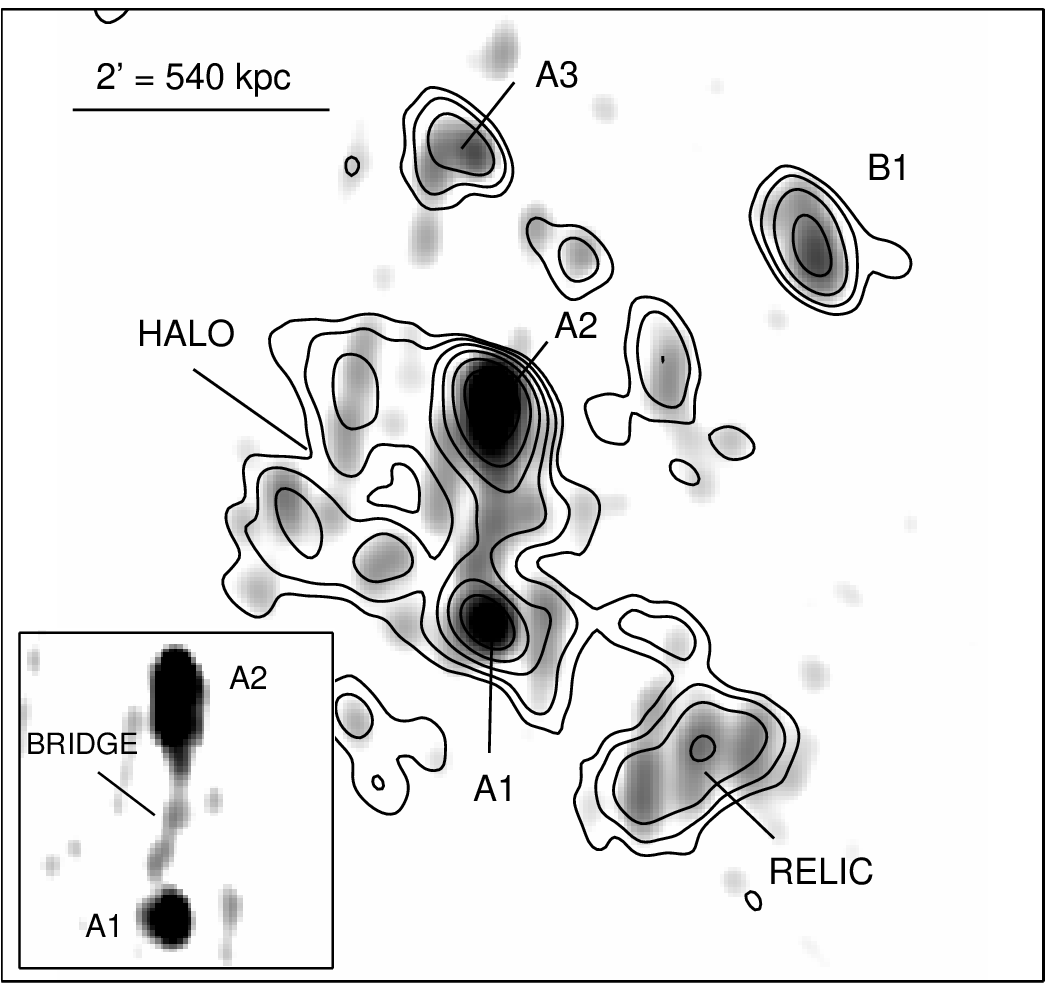}
\hspace{0.2cm}
\includegraphics[scale=0.34]{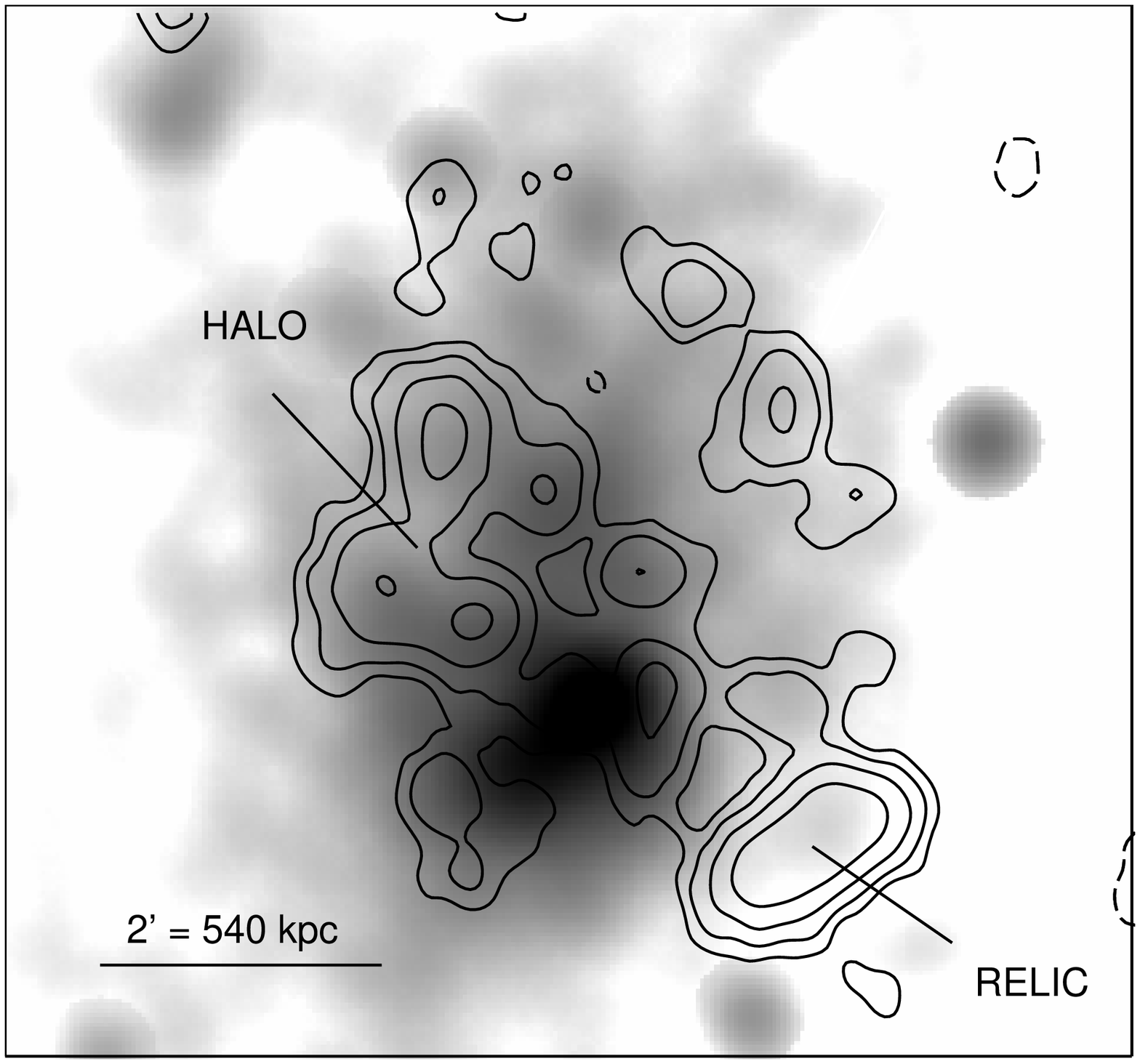}
\caption{\footnotesize {\bf A\,1300}. 
{\it Left:} {\em GMRT} 325 MHz radio contours 
(beam 28$^{\prime\prime}\times18.^{\prime\prime}$), 
overlaid on the full-resolution image (beam 14$^{\prime\prime}\times 9^{\prime\prime}$). 
Contours start at $+3\sigma=$1.8 mJy/beam
and then scale by a factor of 2. Dashed contours correspond to the
$-3\sigma$ level. Individual radio galaxies are
labelled following the notation in Reid et al. (1999). 
{\it Right:} Radio contours after subtraction of the discrete radio
sources, overlaid on the
smoothed {\em Chandra} image. The radio beam is 28$^{\prime\prime}\times28^{\prime\prime}$.
Contours start at $+3\sigma=$1.5 mJy/beam and then scale by a
factor of 2. The $-3\sigma$ level is shown as dashed contours.
}
\label{fig:a1300}
\end{figure*}

A\,1300 is a complex merging cluster with a global temperature 
of $\sim$11 keV, which places the system among the hottest 
clusters known. It hosts a radio halo and a candidate relic \citep{reid99}.

Fig.~\ref{fig:a1300} (left) shows the 325 MHz low-resolution 
contours overlaid on the full resolution image, displayed in grey 
scale to highlight the position and morphology of the radio galaxies 
embedded in the diffuse emission (the sources are labelled following 
the notation in Reid et al. 1999). A faint {\em bridge} of
emission connects the head-tail A2 and point
source A1 (see inset). It is not clear whether such
feature is physically related to A2 or is part of the underlying 
radio halo.
The right panel shows the residual diffuse emission after subtraction 
of the individual radio galaxies (the {\em bridge} was also
subtracted), overlaid on the {\em Chandra} image. 
The radio halo brightness distribution is fairly uniform and
there is little gradient going from the central part to the periphery. 
Its shape is similar to the higher frequency images, with the emission 
extending mainly E of  A1 and A2. The size is $\sim$890 kpc,
comparable to the extent reported in Reid et al. The candidate relic, too, is
similar in shape, extent ($\sim$450 kpc) and brightness
distribution to the earlier images. 

The flux density of the halo at 325 MHz is  
$S_{\rm 325 MHz}$= 130$\pm$7 mJy. 
Reid et al. report $S_{\rm 1.3 \, GHz}$= 10 mJy with {\em ATCA},  while 
the halo has $S_{\rm 1.4 \, GHz}$= 20 mJy on the NVSS .
The spectral index of the halo is $\alpha$=1.3 using 
NVSS and $\alpha$=1.8 using the {\em ATCA} value. 
The candidate relic has
$S_{\rm 325 MHz}$=75$\pm$4 mJy. Its spectral index is
$\alpha$=0.9 between 325 and 843 MHz and  
$\alpha$=1.3 in the 1.3-2.4 GHz range.

Fig.~\ref{fig:a1300} shows that the halo extends mainly NE of 
the X-ray peak, while the
candidate relic is located in the SW periphery of the detected 
X-ray emission. On the basis of its morphology, location 
and steep spectral index, we support the interpretation that this
source is a relic.

\subsection{A\,1758N}

A\,1758N is part of a well-known pair of two merging clusters,
A\,1758N and A\,1758S,
both characterised by very complex X-ray morphology, with several
clumps of emission, as typical of ongoing mergers \citep{kempner04}.
A\,1758N shows more extreme X-ray properties, with higher X-ray
luminosity and temperature compared to A\,1758S. It is also more 
interesting in the radio band, hosting a spectacular diffuse radio source
visible both on the NVSS and WENSS \citep{KS01}. Deep {\em VLA} 
observations at 1.4 GHz were presented in \cite{giovannini09}.

\begin{figure*}[t!]
\centering
\includegraphics[scale=0.34]{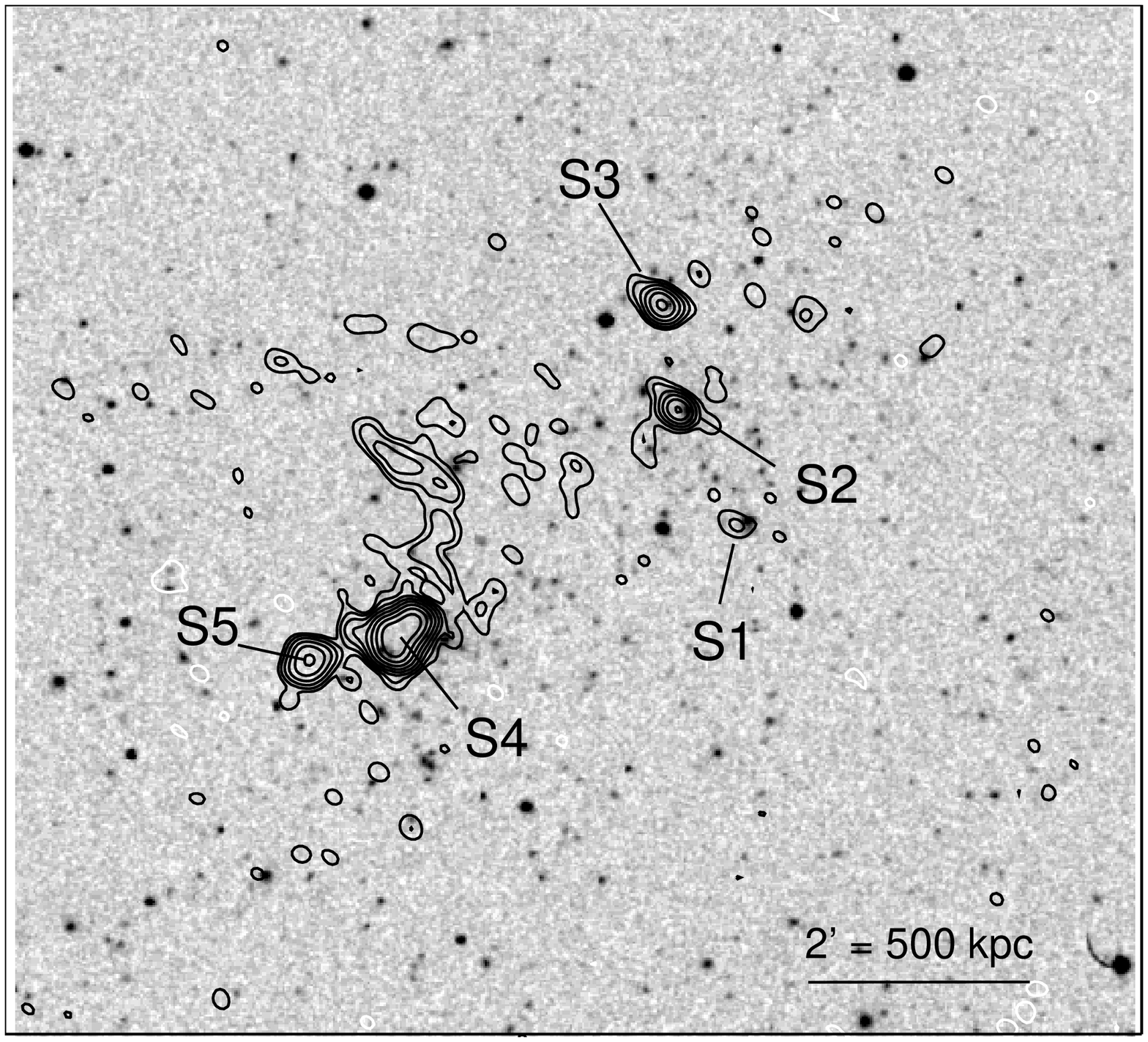}
\hspace{0.2cm}
\includegraphics[scale=0.34]{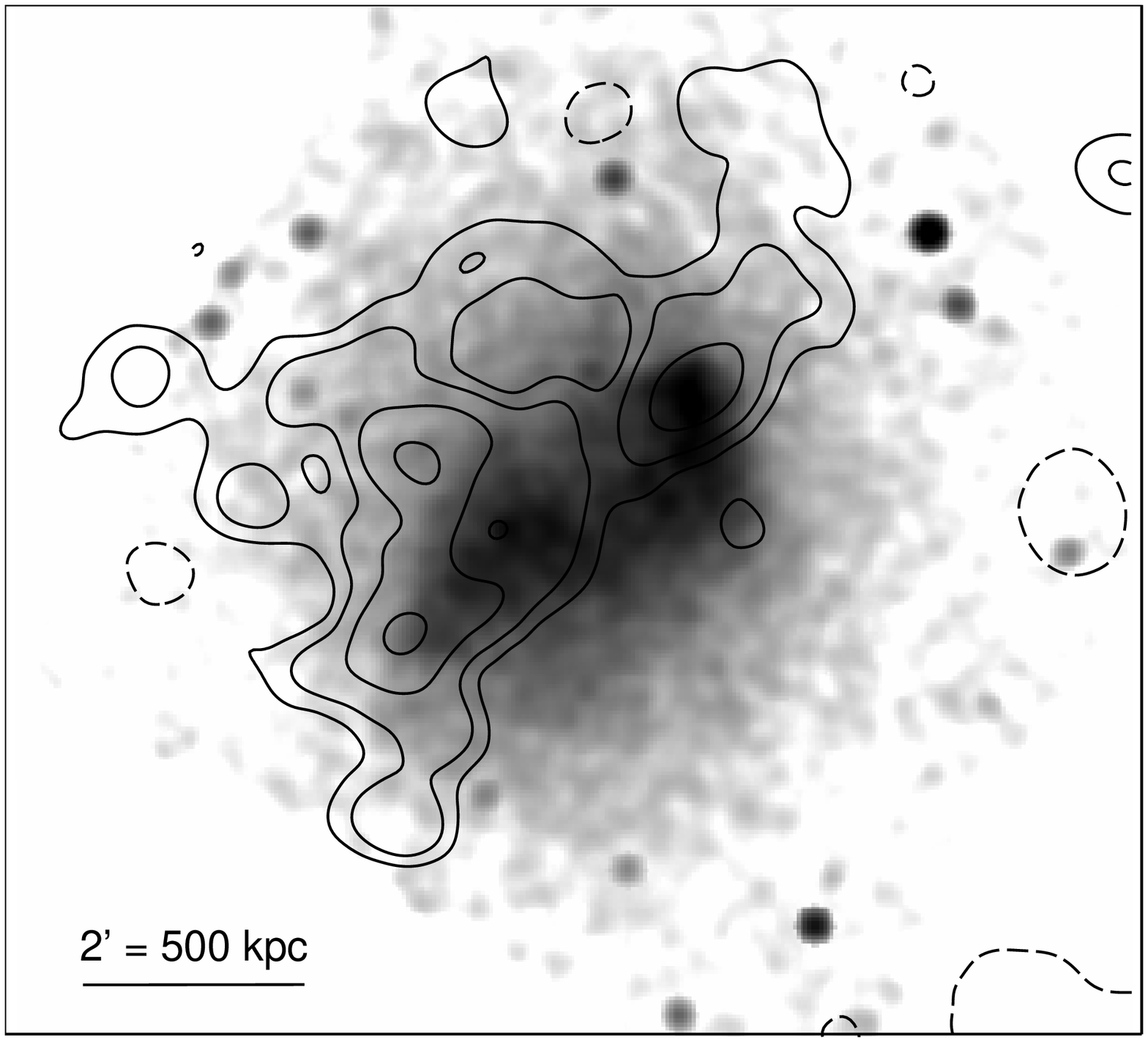}
\caption{\footnotesize {\bf A\,1758}. 
{\it Left:} {\em GMRT} 325 MHz contours at full resolution 
(11$^{\prime\prime}\times8^{\prime\prime}$),
overlaid on the POSS-2 optical image. Contours start at +4$\sigma$=0.4
mJy/beam and then scale by a factor
of 2. Individual radio galaxies are
labelled from S1 to S5. {\it Right:} Radio contours 
after subtraction of the discrete radio sources (S1 to S5), overlaid 
on the smoothed {\em Chandra} image. The radio beam is
35$^{\prime\prime}\times35^{\prime\prime}$. Contours start at
$+3\sigma=$1.2 mJy/beam and then scale by a
factor of 2. The $-3\sigma$ level is shown as dashed contours.
}
\label{fig:a1758}
\end{figure*}

The {\em GMRT} 325 MHz images of A\,1758N are shown in Fig. \ref{fig:a1758}.
On the left, the full resolution contours are overlaid on  
the optical image. The right panel shows the overlay of the 
residual diffuse emission (after subtraction of the discrete sources S1
to S5) on the {\em Chandra} X-ray image. 
The diffuse radio source is filamentary and occupies mainly 
the central and NE region of the cluster.
Its flux density at 325 MHz is 146$\pm$7 mJy and the total 
size is $\sim$ 1.5 Mpc. 

The overall shape and size are in
partial agreement with the {\em VLA} 1.4 GHz image in Giovannini et 
al. (2009), who classify the radio emission as a radio halo
and a double peripheral relic. Indeed the morphology of this source is 
very peculiar, however the spatial coincidence between the radio and 
X-ray emission
(Fig.~\ref{fig:a1758}) suggests that the source is centrally
 located. Hence, here it is classified as a radio halo.
Its spectral index, using the 1.4 GHz flux in Giovannini et al. (2009), is 
$\alpha=1.5$.

\begin{figure*}[t!]
\centering
\includegraphics[scale=0.34]{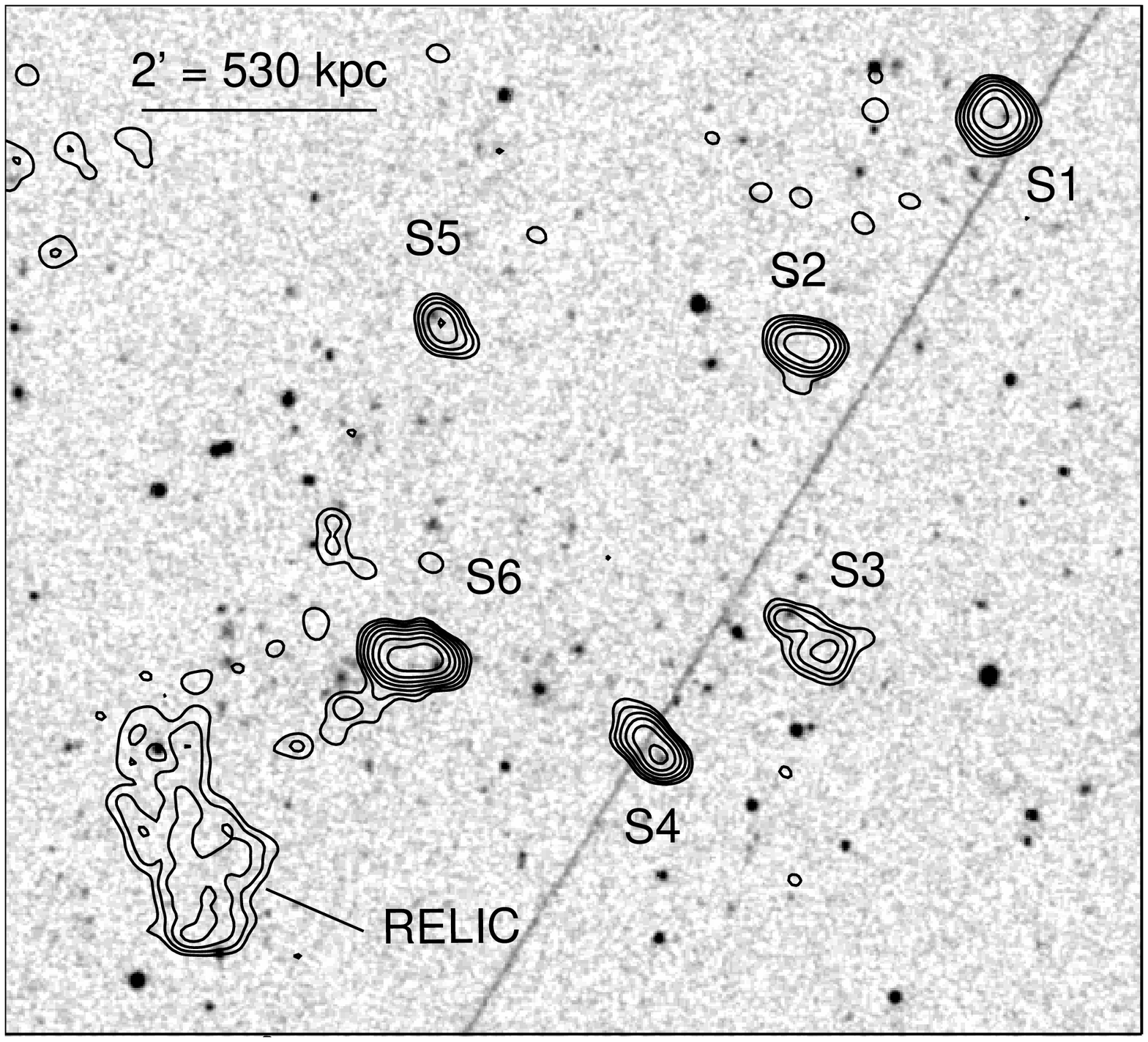}
\hspace{0.2cm}
\includegraphics[scale=0.34]{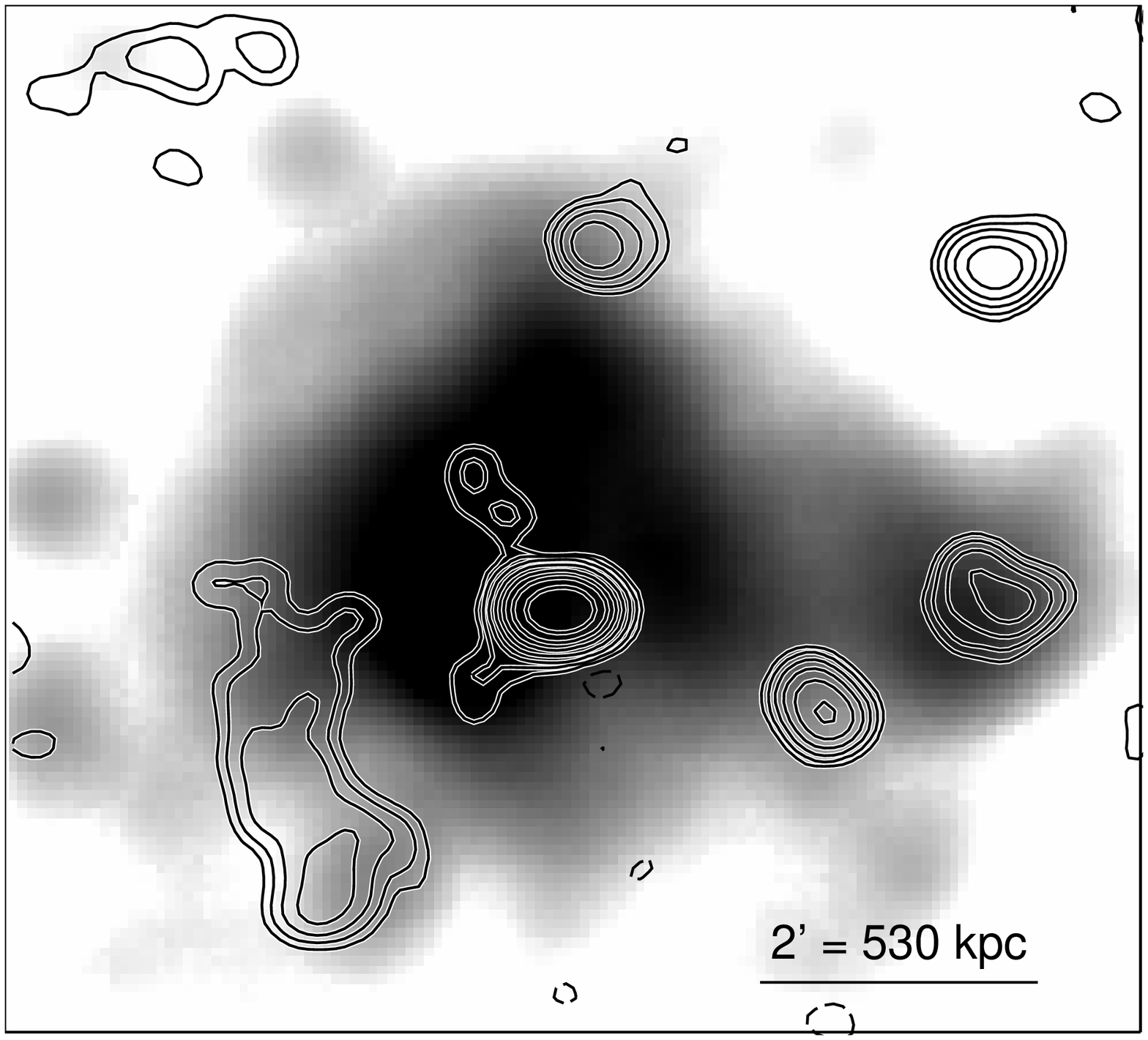}
\caption{\footnotesize {\bf A\,781}. 
{\it Left:} {\em GMRT} 325 MHz contours at full resolution (12$^{\prime\prime}\times9^{\prime\prime}$),
overlaid on the POSS-2 optical image. Contours start at 
$+3\sigma=0.45$ mJy/beam and then scale by a factor
of 2. Individual radio galaxies are
labelled from S1 to S6. {\it Right:} 325 MHz contours at the
resolution of 19$^{\prime\prime}\times 17^{\prime\prime}$,
overlaid on the smoothed {\em Chandra} image. 
Contours start at $+3\sigma=$0.8 mJy/beam and then scale by a
factor of 2. The $-3\sigma$ level is shown as dashed contours.
}
\label{fig:a781}
\end{figure*}

\subsection{A\,781}

A\,781 (z=0.298) has complex X-ray emission, with multiple 
peaks in the central region and a secondary SE condensation at 
$\sim 7^{\prime}$, associated with the galaxy cluster
CXOU J\,092053+302800 (z=0.291). The {\em GMRT} 610 MHz 
observations 
revealed a diffuse source at $\sim$3$^{\prime}$ SE the cluster centre, 
which was classified as
candidate radio relic (Venturi et al. 2008).
The new 325 MHz images are presented in Fig.~\ref{fig:a781}. Beyond
the individual sources (S1 to S6), the candidate relic is the most
striking feature in the field. The source is 
much more extended than at 610 MHz, with a total size of 
$\sim$700 kpc.
The brightness is peaked in the southernmost part of the source, 
which is also edge-brightened, in agreement with the 610 MHz images. 
The flux density of the candidate relic is S$_{\rm 325~MHz}$=93$\pm$5
mJy, fairly consistent at all resolutions.  Flux density measurements at 1.4
GHz from NVSS and archival {\em VLA} images \citep{venturi11b}
provide S$_{\rm 1.4~GHz} \sim$15 mJy. The resulting spectral index is 
$\alpha=$1.25. The nature of this source remains uncertain. Its properties -
morphology, size, peripheral location and steep spectrum - suggest
that it might be indeed a radio relic. However, the 
edge-brightening in the southern region of the source
is unusual for a relic. 

To check for a possible extension of the relic towards the cluster
centre, and/or for a central radio halo undetected at 610 MHz, the
A\,781 field was imaged at low resolution (from $\sim$30$^{\prime
  \prime}$ to $\sim$60$^{\prime \prime}$). Images of the
residual emission after subtraction of the individual radio galaxies
were also produced.  While a $\sim$10 mJy extension of the candidate 
relic is observed in the direction of the double source S6, no large-scale, 
cluster-centre diffuse emission is detected at the sensitivity level 
of these images ($1\sigma \sim 0.25 \div 0.45$ mJy/beam). Flux 
density measurements on the same images suggest residual emission
of the order of 20-30 mJy (including the $\sim$10 mJy extension of the relic) 
over a 1 Mpc-diameter region around the cluster center. 
The total flux density in the individual sources
and in the relic is 526 mJy, and the residual flux density
is $\sim$5\% of this value.
A thorough multi-frequency investigation of A\,781 will be 
presented in Venturi et al. (2011b). 

\section{Summary}

{\em GMRT} observations at 150, 235 and 325 MHz of the 
diffuse radio sources in the {\em GMRT} Radio Halo Survey sample 
(Venturi et al. 2007, 2008) will provide valuable information on the 
radio spectral properties of giant
radio halos and relics \citep{venturi11a}. Here, deep 325 MHz images 
have been shown for the spectacular merging clusters A\,2744, A\,1300, 
and A\,1758N, belonging to the sample and previously known to possess 
impressive radio emission on the Mpc-scale.
Images for A\,781 have been also presented. An extended
radio feature at the periphery of this merging system was classified 
as a candidate relic at 610 MHz. The relic is much more extended in 
the new images at 325 MHz and exhibits a {\it spur} of emission toward 
the cluster centre. Beyond such spur, no further diffuse emission
seems to be present at the cluster center, at least at the sensitivity
level of the 325 MHz images \citep{venturi11b}.

\begin{acknowledgements}
I aknowledge my collaborators T. Venturi, R. Cassano, G. Brunetti,
D. Dallacasa, G. Macario and M. Markevitch. I also thank N. Kantharia, 
R. Athreya and the staff of the {\em GMRT} for their help during 
the observations. {\em GMRT} is run by the National Centre for Radio Astrophysics 
of the TIFR. Support was provided by
NASA through Einstein Postdoctoral
Fellowship PF0-110071 awarded by CXC, which is operated by 
SAO, and by the {\em Chandra} grant AR0-11017X.
\end{acknowledgements}

\bibliographystyle{aa}

\end{document}